# Single-layer behavior and slow carrier density dynamic of twisted graphene bilayer

Lan Meng[1], Yanfeng Zhang[2], Wei Yan[1], Lei Feng[1], Lin He[1,a)], Rui-Fen Dou[1,b)], and Jia-Cai Nie[1]

[1] Department of Physics, Beijing Normal University, Beijing, 100875, People's Republic of China

[2] College of Engineering, Peking University, Beijing, 100871, People's Republic of China

We report scanning tunneling microscopy (STM) and spectroscopy (STS) of twisted graphene bilayer on SiC substrate. For twist angle ~ 4.5° the Dirac point $E_D$ is located about 0.40 eV below the Fermi level $E_F$ due to the electron doping at the graphene/SiC interface. We observed an unexpected result that the local Dirac point around a nanoscaled defect shifts towards the Fermi energy during the STS measurements (with a time scale about 100 seconds). This behavior was attributed to the decoupling between the twisted graphene and the substrate during the measurements, which lowers the carrier density of graphene simultaneously.

Since the pioneering work of Novoselov *et al.* in 2004 [1], graphene has ignited a tremendous outburst of scientific activities in the study of its electronic properties. [2-6] One of the most exciting prospect of this subject is that a generation of electronic devices and circuitry could be made out of graphene. Recent studies indicate that the properties of graphene can be tailored chemically [7-10] and/or structurally [11-16] in many different ways due to its unusual structural and electronic flexibility. This opens doors for an all-graphene circuit in the future.

Epitaxial graphene grown on SiC substrates is a simple and reliable way that offers large scale graphene samples in device fabrication. [11,17,18] However, as-grown epitaxial graphene is intrinsically electron doped due to charge transfer from the SiC substrate. This lowers the Dirac point $E_D$ away from the Fermi energy $E_F$. [19-23] Recently, many approaches have been introduced to remove or compensate the excess charge. Deposition of electron acceptors on top, [24-28] hydrogen intercalation between graphene layers and the substrate, [29] and tunable back-gate electrodes [30,31] are used to tune the carrier concentration of epitaxial graphene.

Here, the topography and local electronic properties of epitaxial bilayer graphene on SiC were studied by scanning tunneling microscopy (STM) and scanning tunneling spectroscopy (STS). For twist angle ~ 4.5° the Dirac point $E_D$ is located about 0.40 eV below the Fermi level $E_F$, which resembles that of single-layer graphene on SiC. The single layer behavior of the twisted graphene indicates electronic decoupling of the surface layer and the sub-layer. A nanoscaled hole of the epitaxial graphene is observed to lower the local Dirac point from 0.40 eV to 0.58 eV below $E_F$. Interestingly, our experiment reveals that the local Dirac point around the defect shifts towards the Fermi energy during the STS measurements, which corresponds to slow dynamics of the carrier density with a time scale ~ 100 seconds. This behavior was attributed to the decoupling between the twisted graphene and the SiC substrate during the STS measurements. By removing the STM tip (also with a time scale ~ 100 seconds), the local Dirac point around the defect returns back to -0.58 eV possibly due to the structure relaxation of the graphene bilayer. The result presented in this paper reveals a method to control the local electronic properties of graphene.

Epitaxial graphene was grown in ultrahigh vacuum (from about $5.0\times 10^{-9}$ to $1.0\times10^{-10}$ Torr) by thermal Si sublimation on hydrogen etched 6H-SiC(000-1).[17,18,32,33] The SiC was heated one hour at about 700° for degassing. The epitaxial graphene was prepared by subsequently annealing the sample at a higher temperature (about 1350°) for 10 min. The thickness of the epitaxial graphene can be controlled by the annealing temperature and time followed by slow cooling to room-temperature. Under these conditions, epitaxial graphene with thicknesses ranging from one to double layers was obtained. The STM system was an ultrahigh vacuum four-probe SPM from UNISOKU. All STM and STS measurements were performed at liquid-nitrogen temperature (78 K) and the images were taken in a constant-current scanning mode. The STM tips used were obtained by chemical corrosion of a tungsten wire. The tunneling conductance, *i.e.*, the $dI/dV$-$V$ curve, was carried out with a standard lock-in technique using a 987 Hz a.c. modulation of the bias voltage.

Figure 1(a) shows a STM image of a typical twisted area of the epitaxial graphene. Periodic protuberance is attributed to the Moire patterns arising from a lattice mismatch (the twist) between the top graphene layer and the underlying layer. Atomically resolved STM topograph of the graphene within the red square is shown in Fig. 1(b). Hexagonal lattices with a periodicity of $a$ = 0.246 nm can be clearly observed. The perfect periodic lattice indicates that the graphene surface is clean and there is no atomic scale defects. The period $D$ of the Moire pattern is about 3.1 nm. The inset of Fig. 1(b) shows the corresponding Fourier transforms. It reveals two sets of peaks arranged in concentric hexagons. The out spots show the graphene reciprocal lattice and the inner hexagon corresponds to the Moire pattern reciprocal lattice. The relative rotation angle in $k$-space between the two hexagons is measured as $\varphi$ ~ 27.7°. Then the twist angle θ in the real space can be estimated by $\theta = (60°-2\varphi) \sim 4.6°$. The value of the twist angle θ is also related to the period of the Moire pattern by $D = a/(2\sin(\theta/2))$. For D ~ 3.1 nm, the twist angle is estimated as 4.5°. Obviously, the above two estimations are consistent. Fig. 1(c) shows the structural model of the Moire pattern. The two misoriented honeycomb lattices are overlaid at an angle of about 4.5°. There is a periodic pattern of points in space at which atoms from the two layers are nearly on top of each other (the structure appears more open in the top view), which is the origin of periodic protuberance in the STM image, as shown in Fig. 1(a).

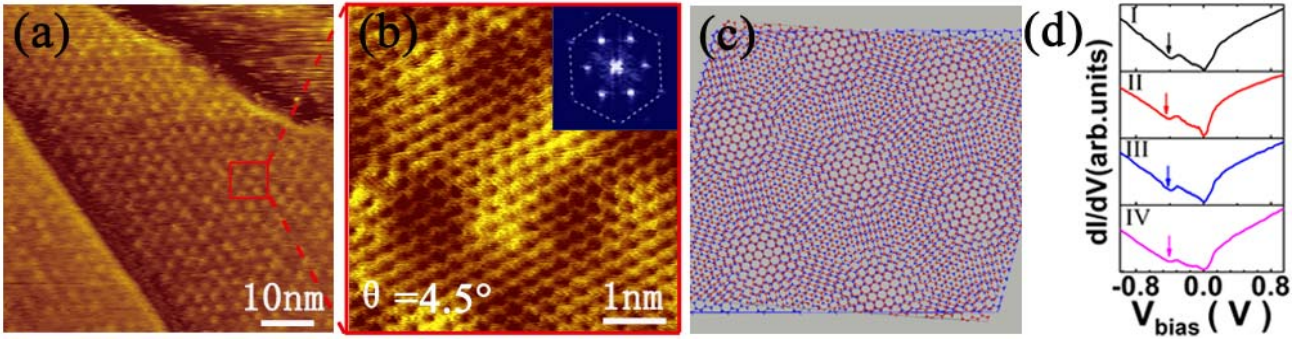

FIG. 1. (Color online) (a) STM image of typical epitaxial graphene on SiC ($V_{sample}$ = -268 mV and I = 0.3 nA). (b) The zoom-in topography of the red square in panel (a) shows atomically resolved STM image of the bilayer graphene ($V_{sample}$ = 173 mV and I = 0.2 nA). The inset is Fourier transform of the main panel showing the contributions from the atomic lattice (out hexagon) and from the Moire pattern (inner hexagon). The relative rotation angle in k-space between the two hexagons is measured as $\varphi \sim 27.7°$. (c) Schematic structural model of two misoriented honeycomb lattices with a twist angle of 4.5°. The structure appears more open in this top view when atoms from the two layers are nearly on top of each other. (d) Four typical dI/dV-V curves obtained on the surface of the epitaxial bilyer graphene. The dI/dV-V curves are measured in the same position one by one. The arrows point to a local minimum ~ -0.40V in the tunneling conductance, representing the Dirac point of graphene.

Figure 1(d) shows four typical dI/dV-V curves on the surface of the epitaxial bilayer graphene. The dI/dV-V curves are completely reproducible. The tunnelling spectrum gives direct access to the local density of states (LDOS) of the surface at the position of the STM tip. A local minimum at about -0.40 V of the tunneling conductance can be attributed to the Dirac point of the graphene.[30,31] The n-type behavior of graphene arises from the charge transfer from the SiC substrate. Our experiment indicates that the Dirac point of the twisted graphene is very stable during the measurements. The result $E_D$ = -0.40 eV is consistent with the well-established result that the Fermi level is located about 0.42 eV above the Dirac point for as-grown monolayer graphene on SiC.[27] Meanwhile, the STS spectra obtained in our experiments consist quite well with that of epitaxial grapnene on SiC substrate reported in literature.[21-23] The single layer behavior of the bilayer graphene is due to the relative rotation between the two layers.[34,35] Very recently, Luican et al. demonstrated that the excitation spectrum of twisted graphene with θ > 3° can still be described by massless Dirac fermions as that of monolayer graphene but with a renormalized Fermi velocity.[34,35] The charge-carrier-concentration n of the twist graphene is estimated as about $2 \times 10^{13}$ cm$^{-2}$ according to

$$n = (E_F - E_D)^2 / (\pi \hbar^2 v_f^2). \quad (1)$$

Here $v_f \sim 7.1 \times 10^5$ m/s is the renormalized Fermi velocity of bilayer graphene with a twist angle of about 4.5°[34] and ℏ the reduced Planck constant.

Figure 2(a) shows a STM topography of the epitaxial graphene with a nanoscaled defect. The profile line of the nanoscaled hole shows that its height is about 1.5 nm. This is much larger than the thickness of bilayer graphene, which may be attributed to the surface defect of SiC substrate. High-resolution STM image (Fig. 2(b)) of the area around the nanoscaled hole indicates that first layer is rotated by about 4.4° relative to the second layers. Recent studies demonstrated that defects of graphene play an vital role in determining the electronic properties of graphene.[36,37] Our experimental results indicate that the local charge-carrier-concentration is enhanced near the nanoscaled hole. The local Dirac point is found to lower from 0.40 eV to 0.58 eV below $E_F$. This result also confirms that the defects of graphene as well as the roughness of the substrate[31] can induce carrier density imhomogeneous of epitaxial graphene. In order to further explore the influence of the nanoscaled hole on the electronic properties of graphene, we carried out controlled experiment of tunneling conductance near the hole. Fig. 2(c) shows typical four dI/dV-V curves aquired in a position around the hole. The dI/dV-V curves from I to

$$\varepsilon_\alpha^2(k) = \frac{\gamma_1^2}{2} + \frac{U^2}{4} + (v^2 + \frac{v_3^2}{2})k^2 + (-1)^\alpha \sqrt{\Gamma} \cdot \quad (2)$$

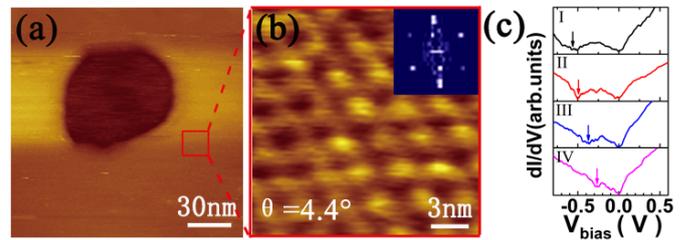

FIG. 2. (Color online) (a) STM topography of epitaxial graphene with a nanoscaled defect ($V_{sample}$ = 0.25 V and I = 0.16 nA). (b) The enlarged view of the red square in panel (a) shows the Moire pattern with a twist angle of 4.4°. The inset is Fourier transform of the main panel. (c) Four typical dI/dV-V curves obtained near the nanoscaled hole of the bilayer graphene. The dI/dV-V curves from I to IV are measured in the same position one by one. The local Dirac point of the graphene is shifted from -0.58 V (I) to -0.25 V (IV) by the STM tip during the measurements.

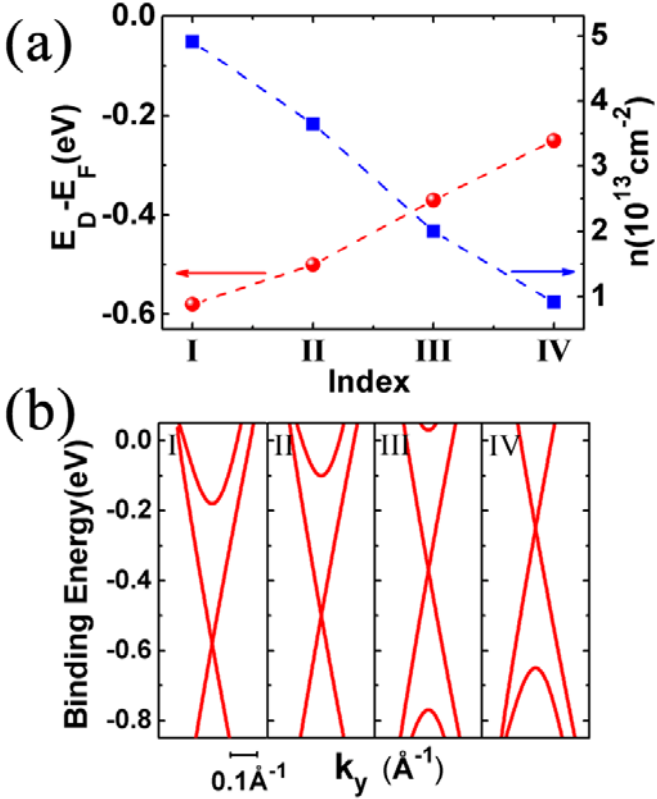

Here the band index $\alpha = 1,2$, $\Gamma = \frac{1}{4}(\gamma_1^2 - v_3^2 k^2)^2 + v^2 k^3 (\gamma_1^2 + U^2 + v_3^2 k^2) + 2\gamma_1 v_3 v^2 k^3 \cos 3\phi$, $k$ is the momentum, $\phi$ the azimuthal angle, $U$ the difference in the onsite Coulomb potentions of the two layers, $\gamma_1$ and $\gamma_3$ the out-of-plane nearest-neighbor and next-nearest-neighbor interaction parameters, and $v_3 = \frac{\sqrt{3} a \gamma_3}{2\hbar}$. Fig. 3(b) shows the calculated tight-binding bands with taking into account the corresponding local charge-carrier-concentration.[39] In our calculation, the renormalized Fermi velocity $v$ of the twisted graphene is also taking into account.[34] For twist angle ~ 4.5°, $v \sim 0.71 v_F \sim 7.1 \times 10^5$ m/s. Here $v_F \sim 1 \times 10^6$ m/s is the Fermi velocity of single layer graphene. Obviously, the local electronic properties of the graphene around the nanoscaled hole are altered during the measurements.

FIG. 3. (Color online) (a) Energy position of the Dirac point during the measurements shown in Fig. 2(c). The right Y-axis shows the carrier concentration calculated according to Eq. (1). (b) Bands of the bilayer graphene near the Fermi level calculated within a tight-binding model. During the measurements, the local Dirac point shifts from -0.58eV to -0.25eV from panel I to IV due to the variation of carrier density.

IV are measured one by one. Surprisingly, the local Dirac point (pointed out by the arrows) shifts toward the Fermi energy. After four STS measurements and by removing the STM tip away from the hole defect, we observed that the local Dirac point returns back to the original position (-0.58 eV) when we carried out a STS measurement subsequently. Then the Dirac point could be shifted toward the Fermi level once again during the next STS measurements. To confirm the experimental result, we carried out several tens of $dI/dV$-$V$ measurements and the obtained phenomenon was completly reproducible around the hole defect. This eliminates any local defects induced during the STS measurements as the possible origin. To further confirm this assertion, we carried out STM measurements around the hole defect after the STS measurement and no obvious difference is observed. Our experimental results indicate that the local charge-carrier-concentration is altered during the STS measurements and recover to initial state with a time scale ~ 100 seconds.

Fig. 3(a) summarizes the energy position of the Dirac point shown in Fig. 2(c). The corresponding carrer concentration calculated by Eq. (1) is shown in the left Y-axis of Fig. 3(a). The local carrier density around the hole is lowered about five times during the measurements. The low-lying electronic tight-binding bands of bilayer graphene can be well described by the following energy dispersion relation[11,38]

What is then the origin of the variation of the local charge-carrier-concentration? The relaxation times of electrons in graphene is usually about dozens of femtosecond.[40,41] In our experiments, it takes about 2 minutes to measure the four $dI/dV$-$V$ curves, which is several orders larger than the relaxation times of electrons in graphene. This eliminates the relaxation of electrons in graphene as the origin of the observed phenomenon. As should be noted, the shift of the Dirac point is reproducible and can only be observed around the nanoscaled hole. Therefore, we can conclude that the hole defect is vital for the observed behavior. For epitaxial graphene grown on SiC, the first carbon layer (interface or buffer-layer) is covalently bonded to the SiC substrate.[29,42,43] The manipulation of the buffer-layer, such as hydrogen intercalation,[29] boron and phosphorus intercalation[42] that break the covalent bonds, can be used to control the charge carriers density of th graphene effectively. Carbon atoms of graphene without covalent bonding to the SiC substrate were also observed in the lattice mismathed regions.[43] In these regions, carbon atoms couple the underlying SiC substrate only through weak van der Waals bonding. In our experiment, carbon atoms in the buffer-layer around the hole defect may possible form weak van der Waals bonding to the underlying SiC substrate. During the STS measurements performed around the hole, the STM tip acts as a top gate electrode that provides an electric field between the tip and sample. As a consequence, the coupling strength between the buffer-layer and the SiC substrate may be easily changed by the electrostatic force between the tip and sample. Then the decrease of local carrier density may arise from weakening of the coupling strength. When we removed the STS tip, the coupling strength is recovered possibly due to the structure relaxation of the graphene bilayer. This is confirmed by remeasuring the Dirac point around the hole, which is lowered back to -0.58 eV.

In conclusion, the topography and local electronic properties of epitaxial bilayer graphene on SiC were studied by STM and STS. Our experiment demonstrates that the twisted graphene with twist angle ~ 4.5° resembles that of single-layer graphene on SiC. A nanoscaled hole of the epitaxial graphene is observed to lower the local Dirac point from 0.40 eV to 0.58 eV below $E_F$, which indicates that defects can induce carrier density inhomogeneous of graphene. Remarkably, our experiment reveals that the local Dirac point around the defect shifts towards the Fermi energy during the STS measurement. It was attributed to the decoupling between the twisted graphene and the SiC substrate. This result reveals a method to control the local electronic properties of graphene.

This work was supported by the National Natural Science Foundation of China (Grant Nos. 10804010, 10974019, 11004010 and 21073003), the Fundamental Research Funds for the Central Universities, and the Ministry of Science and Technology of China (Grants Nos. 2011CB921903).

a): helin@bnu.edu.cn;
b): rfdou@bnu.edu.cn.

1. K. S. Novoselov, A. K. Geim, S. V. Morozov, D. Jiang, Y. Zhang, S. V. Dubonos, I. V. Grigorieva, and A. A. Firsov, Science **306**, 666 (2004).
2. A. K. Geim and K. S. Novoselov, Nat. Mater. **6**, 183 (2007).
3. A. H. Castro Neto, F. Guinea, N. M. R. Peres, K. S. Novoselov, and A. K. Geim, Rev. Mod. Phys. **81**, 109 (2009).
4. S. Das Sarma, S. Adam, E. H. Hwang, and E. Rossi, Rev. Mod. Phys. **83**, 407 (2011).
5. K. S. Novoselov, A. K. Geim, S. V. Morozov, D. Jiang, M. I. Katsnelson, I. V. Grigorieva, S. V. Dubonos, and A. A. Firsov, Nature **438,** 197 (2005).
6. Y. B. Zhang , Y. W. Tan , H. L. Stormer, and P. Kim, Nature **438** , 201 (2005).
7. T. Ohta, A. Bostwick, T. Seyller, K. Horn, and E. Rotenberg, Science **313**, 951 (2006).
8. B. Uchoa, C.-Y. Lin, and A. H. Castro Neto, Phys. Rev. B **77**, 035420 (2008).
9. R. Balog, B. Jorgensen, L. Nilsson, M. Andersen, E. Rienks, M. Bianchi, M. Fanetti, E. Lægsgaard, A. Baraldi, S. Lizzit, Z. Sljivancanin, F. Besenbacher, B. Hammer, T. G. Pedersen, P. Hofmann, and L. Hornekær, Nature Mater. **9,** 315 (2010).
10. F. Schedin, A. K. Geim, S. V. Morozov, E. W. Hill, P. Blake, M. I. Katsnelson, and K. S. Novoselov, Nature Mater. **6,** 652 (2007).
11. C. Berger, Z. Song, X. B. Li, X. S. Wu, N. Brown, C. Naud, D. Mayou, T. B. Li, J. Hass, A. N. Marchenkov, E. H. Conrad, P. N. First, and W. A. de Heer, Science **312**, 1191 (2006).
12. M. Y. Han, B. Ozyilmaz, Y. Zhang, and P. Kim, Phys. Rev. Lett. **98**, 206805 (2007).
13. V. M. Pereira and A. H. Castro Neto, Phys. Rev. Lett. **103**, 046801 (2009).
14. F. Guinea, M. I. Katsnelson, and A. K. Geim, Nature Phys. **6,** 30 (2010).
15. N. Levy, S. A. Burke, K. L. Meaker, M. Panlasigui, A. Zettl, F. Guinea, A. H. Castro Neto, M. F. Crommie, Science **329**, 544 (2010).
16. H. Yan, Y. Sun, L. He, J. C. Nie, and M. H. W. Chan, Phys. Rev. B **85**, 035422 (2012).
17. C. Riedl, U. Starke, J. Bernhardt, M. Franke, and K. Heinz, Phys. Rev. B **76**, 245406 (2007).
18. K. V. Emtsev, A. Bostwick, K. Horn, J. Jobst, G. L. Kellogg, L. Ley, J. L. McChesney, T. Ohta, S. A. Reshanov, J. Rohrl, E. Rotenberg, A. K. Schmid, D. Waldmann, H. B. Weber, and T. Seyller, Nature Mater. **8,** 203 (2009).
19. P. Lauffer, K. V. Emtsev, R. Graupner, Th. Seyller, and L. Ley, Phys. Rev. B **77**, 155426 (2008).
20. H. T. Zhou, J. H. Mao, G. Li, Y. L. Wang, X. L. Feng, S. X. Du, Mullen, and H. -J. Gao, Appl. Phys. Lett. **99**, 153101 (2011).
21. J. Červenka, K.van de Ruit, and C. F. J. Flipse, Phys. Rev. B **81**, 205403 (2010).
22. J. Choi, H. Lee, and S. Kim, J. P. Chem. C **114**,13344 (2010).
23. V. B. Brar, Y. B. Zhang, and Y. Yayon, Appl. Phys. Lett. **91**, 122102 (2007).
24. S. Y. Zhou, D. A. Siegel, A. V. Fedorov, and A. Lanzara, Phys. Rev. Lett. **101**, 086402 (2008).
25. B. Premlal, M. Cranney, F. Vonau, D. Aubel, D. Casterman, M. M. De Souza, and L. Simon. Appl. Phys. Lett. **94**, 263115 (2009).
26. I. Gierz, C. Riedl, U. Starke, C. R. Ast, and K. Kern, Nano lett. **8**, 4603 (2008).
27. C. Coletti, C. Riedl, D. S. Lee, B. Krauss, L. Patthey, K. von Klitzing, J. H. Smet, and U. Starke, Phys. Rev. B **81**, 235401 (2010).
28. W. Chen, S. Chen, D. C. Qi, X. Y. Gao, and A. T. S. Wee , J. Am. Chem. Soc. **129**, 10418 (2007).
29. C. Riedl, C. Coletti, T. Iwasaki, A. A. Zakharov, and U. Starke, Phys. Rev. Lett. **103**, 246804 (2009).
30. Y. Zhang, V. W. Brar, F. Wang, C. Girit, Y. Yayon, M. Panlasigui, A. Zettl, and M. F. Crommie, Nature Phys. **4,** 627 (2008).
31. R. Decker, Y. Wang, V. W. Brar, W. Regan, H.-Z. Tsai, Q. Wu, W. Gannett, A. Zettl, and M. F. Crommie, Nano Lett. **11,** 2291 (2011).
32. H. Huang, W. Chen, S. Chen, and A. T. S. Wee, ACS Nano **2**, 2513 (2008).
33. C. Berger, Z. M. Song, T. B. Li, X. B. Li, A. Y. Ogbazghi, R. Feng, Z. T. Dai, A. N. Marchenkov, E. H. Conrad, P. N. First, and W. A. de Heer, J. Phys. Chem. B **108**, 19912 (2004).
34. A. Luican, G. H. Li, A. Reina, J. Kong, R. R. Nair, K. S. Novoselov, A. K. Geim, and E. Y. Andrei, Phys. Rev. Lett. **106**, 126802 (2011).
35. A. H. MacDonald and R. Bistritzer, Nature **474**, 453 (2011).
36. J. Lahiri, Y. Lin, P. Bozkurt, I. I. Oleynik, and M. Batzill, Nature Nanotech. **5**, 326 (2010).
37. M. M. Ugeda, D. Fernandez-Torre, I. Brihuega, P. Pou, A. J. Martinez-Galera, R. Perez, and J. M. Gomez-Rodriguez, Phys. Rev. Lett. **107**, 116803 (2011).
38. E. McCann and V. I. Fal'ko, Phys. Rev. Lett. **96**, 086805 (2006).
39. The tight-binding bands are calculated with U = 0, $\gamma_1$ = 0.4 eV, $\gamma_3$ = 0.12 eV , $\phi$ = 90°, $a$ = 0.246 nm, and $v$ = 7.1×10$^5$ m/s.
40. M. Breusing, C. Ropers, and T. Elsaesser, Phys. Rev. Lett. **102**, 086809 (2009).
41. E. Sano, Appl. Phys. Express **4**, 085101(2011).
42. F.Varchon, R. Feng, J. Hass, X. Li, B. N. Nguyen, C. Naud, P. Mallet, J.-Y. Veuillen, C. Berger, E. H. Conrad, and L. Magaud, Phys. Rev. Lett. 99, 126805 (2007).
43. S. Kim, J. Ihm, H. J. Choi, and Y.-W. Son, Phys. Rev. Lett. 100, 176802 (2008).